# Predictive pion-quark BCS relation and Thornber-Feynman high-$T_c$ gap

M.D. Scadron[a]
B.A. Green[b]
[a]*Physics Department, University of Arizona, Tucson, AZ 85721*
[b]*11912 N. Centaurus Pl., Tucson AZ 85737*

A pion-quark pairing temperature is defined by a BCS-like relation identified from a quark-level Goldberger-Treiman relation with a Nambu scalar mass "gap parameter" taken in the low-mass limit. This intuitive relation predicts the associated "experimental" lattice-gauge pairing temperature. The opposite high-mass limit predicts the sigma mass, and notably has a predictive analog in high-$T_c$ superconductivity in the stable nondispersive energy gap as defined by Thornber-Feynman polaron dynamics.

A predictive analog pion-quark BCS relation with a Nambu-mass gap parameter has a further parallel in the nondispersive high-$T_c$ superconductivity gap as defined from Thornber-Feynman (TF) polaron dynamics [1]. This BCS-like relation for the pion $\bar{q}q$ binding energy and pair-breaking "temperature", as defined from a quark-level Goldberger-Treiman relation (GTR) in the low-mass limit [1], predicts this latest calculated temperature as follows. The GTR for the pion-nucleon coupling constant $g_{\pi NN}$ is $g_{\pi NN}=m_N g_A/f_\pi$ in terms of the nucleon mass $m_N$ (939 MeV), pion decay constant in the chiral pairing limit $f_\pi \sim 90$ MeV, and axial current form factor $g_A=1$ for structureless quarks in the constituent quark model [2]. For the constituent quark this relation becomes $g_{\pi qq}=m_q/f_\pi$ where $m_q \approx 939$ MeV/3$\approx$313 MeV and $g_A$ is unity for structureless quarks, so an "experimental" two-flavor quark-level GTR is

$$g_{\pi qq}=m_q/f_\pi \sim 313 \text{ MeV}/90 \text{ MeV} \approx 3.5. \tag{1}$$

Noting that the righthand side of Eq. 1 numerically approximates the BCS ratio, we cast it into "BCS form" by multiplying the numerator and denominator by two, i.e.,

$$g_{\pi qq} \approx 2m_q/2f_\pi \sim 3.5. \tag{2}$$

The numerator $2m_q$ is recognized as the scalar meson mass identified from the Nambu scalar mass (Eq. A-3) in the Appendix as an analog superconducting gap $2\Delta$. From this approximate BCS ratio $2\Delta/k_B T_c \sim 3.5$ in Eq. 2 for $T_c \sim 3\text{-}4$ K we identify the numerator ($2m_q$) with a pair binding energy $2\Delta$, where the constituent quark masses are largely taken up in the binding as with the pion [3]. This further suggests the BCS denominator $2f_\pi$ is an intuitive $\bar{q}q$ pairing or chiral restoration "temperature" $T_c$, i.e., $T_c=2f_\pi \sim 180$ MeV [4], which in fact is compatible with the recent "experimental" value of 173±8 MeV obtained from a computer "lattice" calculation for two quark flavors [5]. Notably this intuitive $T_c$ determination from the quark-level GTR complements the earlier ones in Ref. 4 in the framework of chiral symmetry breaking.

This Nambu scalar mass-pion BCS connection in the low-mass limit ($2m_q \rightarrow 2\Delta$, $2f_\pi \rightarrow T_c$) is further supported by independent calculations of the BCS ratio and $g_{\pi qq}$ as analytic expressions that are similar in both magnitude and form. Namely for acoustical phonons the gap ratio becomes [1]

$$2\Delta/k_B T_c = 2\pi e^{-\gamma} \approx 3.528, \qquad (3)$$

where $\gamma$ is the Euler constant (0.5772), and for the pion-quark coupling constant with color number $N_c=3$ we find [6,4]

$$g_{\pi qq} \approx 2\pi/N_c^{1/2} = 2\pi/3^{1/2} \approx 3.628. \qquad (4)$$

Note this numerical nearness of Eqs. 3 and 4 to the BCS ratio in Eqs. 1,2.

The above identification of the double quark mass $2m_q$ in Eq. 2 with the pion $\bar{q}q$ pairing energy $2\Delta$ is reinforced by theoretical and experimental determinations of the pion charge radius $r_\pi$ from a "fused" quark pair that accords with the tight-binding massless Nambu-Goldstone pion implicit in the GTR, Eq. 1. Hence $r_\pi$ is specified by a *single* quark mass suggestive of the fused pair as $r_\pi = \hbar c/m_q = 197.3$ MeV-fm/313 MeV $\approx 0.63$ fm ($\hbar c=197.3$ MeV-fm) [7] from the "common sense" quark mass 313 MeV$=m_N/3$ as expected with three constituent non-strange quarks in a nucleus. Similarly, vector meson dominance (VMD) [8] with rho-meson mass 770 MeV specifies a pion charge radius $6^{1/2}$(197.3 MeV-fm)/770 MeV≈0.63 fm.

Not surprisingly, the numerical closeness of 3.528 and 3.628 in Eqs. 3 and 4 is not accidental. In both cases the energy scales with the momentum as E=p (in units c=1), but for different reasons. For the acoustic-phonon BCS coupling in Eq. 3 the linear scaling is in the low-energy limit due to the large Fermi surface (FS), which follows from the relation $\Delta E=(p/m)\Delta p$ with constant of proportionality p/m defined from energy-momentum changes $\Delta E$ and $\Delta p$. Hence p/m is approximately constant on the FS because the changes $\Delta p/p$ are small. Whereas in Eq. 4 the linear scaling is due to tight binding, where the fused pion mass vanishes in the relativistic energy-mass relation $E^2=p^2+m^2$ as the mass converts to binding energy, so E is approximately p in this (tight binding) limit.

The Nambu scalar mass gap parameter Eq. A-3 in the opposite high-mass limit predicts the sigma mass as shown in the Appendix. The further analog in the flatband high-$T_c$ superconductivity gap $2\Delta_{HTC}$ in this limit, paralleling the long-range low-$T_c$ gap parameter Eq. A-2 [1], is *experimentally* specified by the non-dispersive (massive) energy-constrained TF quasiparticle shift $E_{LO}$ (the longitudinal-optical (LO) phonon mode energy) [9]. This mobile, stable, gapped quasiparticle in paralleling the high-$T_c$ particle arises from nonadiabatic (conserved) internal polaron dynamics in such low-symmetry Frohlich media [10] as defined by a "universal" TF mobility [1] (e.g., in photoconducting alkali halides and related transition-metal oxides [12]). Accordingly the optimum, constrained high-$T_c$ gap analogous to Eq. A-2 is empirically

$$2\Delta_{HTC} = E_{LO}, \quad (5a)$$

where in the cuprates this gap as defined by angle-resolved photoemission spectroscopy (ARPES) data in fact equals the resonance-softened in-plane Cu-O mode energy $E_{LO}^R$ in satisfying Eq. 5a directly or from combined cuprate data [1], so the above TF gap relation becomes

$$2\Delta_{HTC} = E_{LO}^R. \quad (5b)$$

In particular for optimum-doped $Bi_2Sr_2CaCu_3O_8$ (Bi-2212) this ARPES energy shift $2\Delta_{SC}$ in Fig. 1(a) is 74 meV (1 meV= $10^{-3}$ eV), where combining with resonance-softened $La_{2-x}Sr_xCuO_4$ (La-214) Cu-O mode data $E_{LO}^R \approx 70$ meV from Table I. Whereas for $YBa_2Cu_3O_7$ (Y-123) the ARPES gap Eq. 5b is predicted by this Y-123 data alone, with the shift $2\Delta_{HTC}$ and $E_{LO}^R$ from Table I and Fig. 1(b) respectively 58 meV and 56 meV [20]. Notably Eq. 5b is further given by a directional resonance coupling of these TF dynamics to the ARPES distribution [1].

Moreover Eq. 5 in addition to a Nambu-sigma mass analog constitutes an empirical "slowing" relation, where the simple energy and coupling independence reflects nondispersive, nonlinear inelastic-Frohlich slowing, and the conservation or irreversibility derives from the system asymmetry [1]. Such energetic "bunching" from these nonlocal dynamics, in setting the (TF) polaron apart from that due to Feynman [10], is consistent with the short-range stable normal-state pairing. Similarly, such nonlinear extension of this Frohlich "action" [11b] facilitates its energetic sharing in the many-particle state.

Significantly an extra-high-$T_c$ multi-layer cubic BCS cuprate possibly observed as a minority phase is defined from this empirical TF Nambu-mass analog gap relation Eq. 5b for the non-cubic cuprates, and in particular for the BCS-like/cubic alkali-fullerene $Rb_2CsCo_{60}$ [21]. Hence in the latter where $T_c$=40 K the TF shift of 13 meV from Table I yields the expected near-BCS ratio,

$$2\Delta_{HTC}/T_c k_B = (13 \text{ meV}/40 \text{ K})(11.6 \text{ K/meV}) \approx 3.8, \qquad (6a)$$

paralleling a low-$T_c$ Nambu-analog gap mode in $NbSe_2$ with gap ratio 3.4 in Eq. A-1. Whereas for the (non-cubic) cuprates non-BCS gap ratios are specified, viz., for Bi-2212 with $T_c$=95 K and $2\Delta_{HTC}$=74 meV experimentally from the TF-ARPES relation Eq. 5b, the ratio is

$$2\Delta_{HTC}/T_c k_B = 74 \text{ meV}/95 \text{ K})(11.6 \text{ K/meV}) \approx 9.0 \qquad (6b)$$

($k_B$=1 meV/11.6 K). Similarly for Y-123 where $T_c$=93 K and $2\Delta_{HTC}$=58 meV again given by Eq. 5b a decreased ratio is specified as

$$2\Delta_{HTC}/T_c k_B = 58 \text{ meV}/93 \text{ K})(11.6 \text{ K/meV}) \approx 7.2, \qquad (6c)$$

where this decrease would reflect the relative uniformity of the Y-123 structure. Notably for the purposes of our projected ideal BCS cuprate, the diversity in the optimum high-$T_c$ gap of these structures of 13 meV, 74 meV and 58 meV is *fundamentally* specified by the TF shift Eq. 5. (Albeit the accompanying $T_c$ increases in the cuprates are offset by the large ratios reflecting non-BCS localization associated with the low lattice symmetry).

Added basis for this ideal BCS cuprate comes from such high-$T_c$ superconductivity associated with cubic symmetry *in general* as in optimum $Rb_2CsC_{60}$ above. Hence in $Ba_{0.6}K_{0.4}BiO_3$ with $2\Delta_{HTC}$=70 cm$^{-1}$ and $T_c$=30 K [22] (recall $\hbar c$=197.3 MeV-fm=0.01973 meV-cm) the gap ratio is

$$2\Delta_{HTC}/T_c k_B = (2\pi)(70 \text{ cm}^{-1}/30 \text{ K})(11.6 \text{ K/meV})(0.01973 \text{ meV-cm}) \approx 3.4, \qquad (7a)$$

and in $Rb_3C_{60}$ with $\Delta_{SC}/k_B$=53 K, or $\Delta$=4.6 meV and $T_c$=29.4 K [23] the ratio is

$$2\Delta_{HTC}/T_c k_B = 2(4.6 \text{ meV}/29.4 \text{ K})(11.6 \text{ K/meV}) \approx 3.6. \quad (7b)$$

These Nambu-analog TF gap and BCS ratio specifications from Eqs. 5b, 6 and 7 provide an empirical basis for defining the ideal BCS cuprate, e.g., the gap specifications imply the TF shift is *fundamental* and therefore also holds for the record-high-$T_c$ 164 K multi-layer structure $HgBa_2Ca_2Cu_3O_8$ under applied hydrostatic pressure [24]. Hence taking the resonance LO-mode energy $E_{LO}^R$ in Eq. 5b in this case as ~65 meV, as estimated from an average of the Y-123 and La-214 data, the gap ratio is logically reduced from 7-9 in Bi-2212 and Y-123 to

$$2\Delta_{HTC}/T_c k_B \sim (65 \text{ meV}/164 \text{ K})(11.6 \text{ K/meV}) = 4.6, \quad (8)$$

where this reduction is consistent with symmetry increases toward a cubic BCS structure from the added planes and hydrostatic pressure.

In Ref. 1 a ~250 K infinite-layer cubic cuprate superconductor was projected from this predictive TF gap Eq. 5b plus BCS condition generally observed in cubic high-$T_c$ structures (cf. Eqs. 6a and 7). For such a Bi-2212 base structure this $T_c$ limit from Eq. 6b as a result of added planes and applied pressure is thus

$$T_c = 74 \text{ meV}(11.6 \text{ K/meV})/3.5 = 245 \text{ K}. \quad (9a)$$

Stated in reverse, the BCS ratio limit is reached for the ideal 245 K $T_c$ structure as

$$2\Delta_{HTC}/T_c k_B = (74 \text{ meV}/245 \text{ K})(11.6 \text{ K/meV}) = 3.5. \quad (9b)$$

Notably this $T_c$ is consistent with numerous observations of possible minority-phase superconductivity near this temperature.

Summarizing this Nambu-analog gap ratio progression toward the extra-high-$T_c$ structure, starting with the low-$T_c$ analog NbSe$_2$ with a ratio 3.4 in Eq. A-1, the sequence continues in the optimum high-$T_c$ structures with ratios 3.8, 9.0 and 7.2, all given by the TF gap Eq. 5. Continuing to higher $T_c$ a TF-lattice symmetry gap ratio pattern peaking at 9.0 emerges with increasing $T_c$ as 3.8, 9.0, 7.2, 4.6(?) → 3.5(?), with (?) denoting the TF-BCS-like projections.

In conclusion, the latest calculated pion-quark pairing temperature is predicted by an intuitive BCS relation identified from a quark-level Goldberger-Treiman relation with a Nambu scalar-mass gap parameter taken in the low-mass limit. Whereas in the high-mass limit this parameter predicts the sigma mass, and moreover has an analog in high-$T_c$ superconductivity in the stable localized energy gap as specified by the Thornber-Feynman (TF) polaron with a strongly-constrained shift from conserved dynamics paralleling the high-$T_c$ case. Notably an extra-high-$T_c$ BCS cuprate possibly observed as a minority phase is defined by a high-$T_c$ gap ratio sequence based on these empirical Nambu-analog TF gaps.

**Appendix: Particle physics analog of a low-$T_c$ gap mode**

We outline here how a massive "gap mode" appearing at $T_c$ near $2\Delta$ in the BCS superconductor NbSe$_2$ constitutes a particle physics analog leading in the low-mass limit to the pion pairing BCS relation Eq. 2. A near-BCS gap ratio is defined in NbSe$_2$ from $2\Delta = 17$ cm$^{-1}$ (with $\omega = 2\pi\nu$) and $T_c = 7.2$ K [25],

$$2\Delta/T_c k_B = [(2\pi)17 \text{ cm}^{-1}/7.2 \text{ K}][0.01973 \text{ meV-cm}][11.6 \text{ K/meV}] \approx 3.4, \quad \text{(A-1)}$$

suggesting a nominal global pairing over the Fermi surface. The particle physics connection of this gap mode follows from its identification by Littlewood and Varma with the Nambu scalar mass [26]. This mode arises from a coupling of the superconductivity/gap to an amplitude mode of the charge-density-wave (CDW) (created from optical-phonon induced oscillation of the CDW gap at a local zone face). Hence as a local paired bound state near 2Δ with mass M, or

$$2\Delta \approx M, \qquad (A-2)$$

it has an analog in the Nambu scalar mass [27] identified in Eq. 2, i.e.,

$$2m_q = m_s. \qquad (A-3)$$

This "gap parameter" Eq. A-3 in the high-mass limit predicts the sigma meson mass $m_\sigma$, as also concluded in Ref. 28 from a chiral-breaking analog mode of Nambu and Jona-Lasinio [27], where $m_q$ is the constituent quark mass as verified experimentally by the scalar sigma with mass $m_\sigma \approx 600$ MeV [29]. Independent determination of the quark mass $m_q$ of 325 MeV in the chiral limit from the linear σ model [6] (or simply $m_N/3 \approx 313$ MeV from the 3 quarks in the nucleus) yields 626-650 MeV for $m_\sigma$. Equation A-3 expresses the dispersion-theoretic limit in the dispersion range with squared four-momentum $q^2=4m^2$ to 0, where the quark-anti-quark pair can be considered as "touching" with $q^2=(2m_q)^2$.

A BCS-particle physics analog of Eq. A-2 in the opposite zero-mass limit in the text gives the predictive BCS-like relation for the pion with the $\bar{q}q$ pair considered as "fused", i.e., a tightly-bound pair at $q^2=m_\pi^2=0$, or at the VMD value $r_\pi \approx 0.63$ fm.

|  | Optimum gap, $2\Delta$ (meV) | In-plane LO-phonon energy (meV) $E_{LO}$ | In-plane LO-phonon energy (meV) $E_{LO}^R$ | Critical temperature, $T_c$ (K) |
|---|---|---|---|---|
| $Bi_2Sr_2CaCu_2O_8$ (Bi-2212) | 74 [12] | 80 [15] |  | 95 |
| $La_{1.85}Sr_{0.15}CuO_4$ (La-214) | 15 | 83 [16] | 70 [16] | 40 |
| $YBa_2Cu_3O_7$ (Y-123) | 58[a] [14] | 68[a] [17] | 56[a] [17] | 93 |
| $Rb_2CsC_{60}$ | 12[b] | 13 [18] |  | 40 [19] |

[a] a-axis data in direction of reduced screening across charge stripes.
[b] This gap with the observed $T_c$ of 40 K corresponds to the BCS gap ratio generally observed in these structures.

**Table I**. Optimum gap $2\Delta$ and non-resonance and resonance in-plane LO-phonon energies $E_{LO}$ and $E_{LO}^R$ in the (1,0,0) direction in Bi-2212, La-214 and Y-123 from ARPES and inelastic neutron scattering data, respectively. Note the $2\Delta$-$E_{LO}^R$ resonance match according to the TF polar coupling relation Eq. 5b from such resonance "softening" of the mid-zone Cu-O mode, and similarly in $Rb_2CsC_{60}$ according to Eq. 5a. The small gap in La-214 reflects the short-range order.

**Figure 1**

(a) ARPES distribution of Bi-2212 at ($\pi$,0) momentum location on the Fermi surface that defines the optimum energy gap $\Delta$ for the normal and superconducting states (from Ref. 13). (b) High-density a-axis superconducting peak defining the gap of Y-123 at ($\pi$,0), with the corresponding Bi-2212 peak (from Ref. 14).

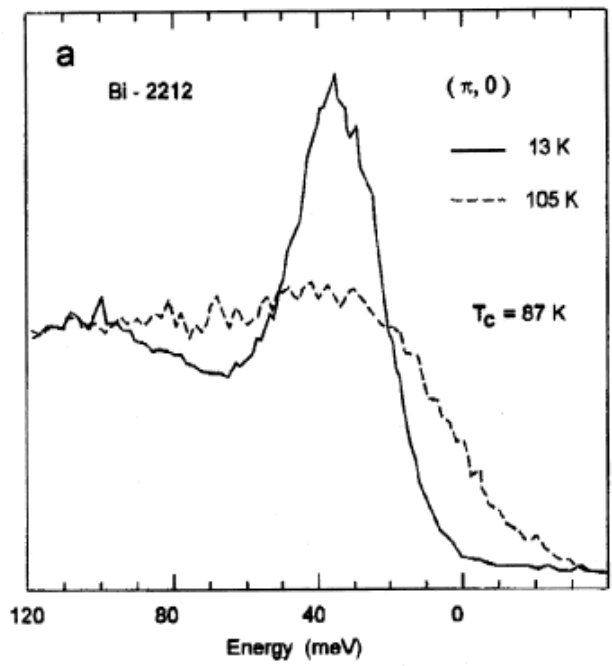
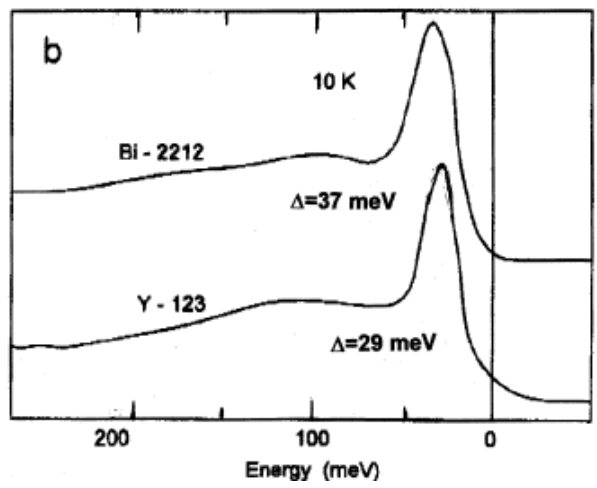